\documentclass[aps,twocolumn,prb,superscriptaddress,amsmath,tightenlines]{revtex4}
\usepackage[dvipdfmx]{graphicx}
\usepackage{graphicx}
\usepackage{epsfig}

\newcommand{\rw}{\rightarrow}
\newcommand{\Rw}{\Rightarrow}

\begin{document}

\title{Generation of all-to-all connections in a two-dimensional qubit array 
with two-body interactions}

\author{Tetsufumi Tanamoto}

\affiliation{Department of Information and Electronic Engineering, Teikyo University,
1-1 Toyosatodai, Utsunomiya 320-8551, Japan}

\begin{abstract}
All-to-all connections are required in general quantum annealing machines to solve various combinatorial optimization problems.
The  Lechner, Hauke, and Zoller (LHZ) method, which is used to realize 
the all-to-all connections, requires many-body interactions in locally connected qubits. 
Because most of the qubit interactions are two-body interactions, 
Lechner also proposed the construction of each four-body interaction 
by six controlled-NOT (CNOT) gates between two qubits. 
However, it is difficult to construct many CNOT gates.
Herein, we show more concrete sequences to 
produce four-body and three-body interactions based on a two-dimensional solid-state qubit system. 
We show that the number of operations needed to construct the many-body interactions can be reduced
using appropriate pulse sequences. These findings will help reduce quantum computation costs for solving combinatorial problems.
\end{abstract}
\maketitle

\section{Introduction}
The progress of artificial intelligence (AI) in science and technology is leading to significant changes in society.
Faster solving of combinatorial optimization problems is a prerequisite condition for 
efficient development of AI algorithms such as deep-learning machine algorithms. 
The quantum annealing machine (QAM) is expected to efficiently solve the combinatorial
optimization problems in a shorter time than 
is possible with classical annealing methods~\cite{Nishimori,Nishimori2,Farhi,Brends,Weber,Ohzeki,Tanaka,Mukai,Maezawa,Kawabata}. 
Nishimori {\it et al.} developed the theoretical foundation of the QAM,
\cite{Nishimori,Nishimori2,Farhi}, and QAMs based on superconducting circuits are 
widely used~\cite{Dwave1,Dwave2}.
In a QAM, NP-hard problems, such as the traveling salesman problem,
can be mapped to problems in finding the ground states of the Ising Hamiltonian,
 expressed by~\cite{Lucas}
$
H=\sum_{i<j} J_{ij} s_i^z s_j^z 
+ \sum_i h_i s_i^z, 
$
where the variable $s_i^z$ is a classical bit of two values ($s_i^z=\pm 1$).
The first term denotes the interaction, with a coupling constant $J_{ij}$, 
and the second term denotes the Zeeman energy with an applied magnetic field $h_i$.
For a QAM, a tunneling term is added and the Hamiltonian is given by
$ 
H=\sum_{i<j} J_{ij} Z_i Z_j 
+ \sum_i [h_i Z_i + \Delta_i (t) X_i],
$
and the variables are expressed by Pauli matrices given by
$
X=
\left(
\begin{array}{cc}
0 & 1 \\
1 & 0 \\
\end{array}
\right)$, 
%Y=
%\left(
%\begin{array}{cc}
%0 & -i \\
%i & 0 \\
%\end{array}
%
and 
$Z=
\left(
\begin{array}{cc}
1 & 0 \\
0 & -1 \\
\end{array}
\right).
$
%%%%%% transformation of ZZ  %%%%%%%%
The tunneling term is controlled such that it disappears at the end of the calculation, given by
$
\Delta(t\rightarrow \infty) \rightarrow 0.
$
To solve many combinatorial problems, all connections between two cells are required.
By contrast, 
interactions between solid-state qubits are limited 
to the nearest or next-nearest interactions.
Choi introduced the minor embedding method to 
solve this problem in the D-Wave superconducting circuit structure.\cite{ME1,ME2} 
Lechner, Hauke, and Zoller (LHZ) proposed a novel method 
of realizing all connections by introducing a logical spin.\cite{LHZ}
Albash {\it et al.} compared the minor embedding and LHZ methods in terms of error tolerance and concluded that the minor embedding method is more error-tolerant 
than the LHZ method.\cite{Lidar} 
However, the best method is decided according to the system, and 
it is better that both methods should be equally investigated.
Here we would like to investigate the LHZ method theoretically.
One of the key challenges when using the LHZ method is to 
construct the four-body interactions.
Kerr nonlinearity based on Josephson parametric oscillators is one of the promising candidates 
proposed for realizing the LHZ scheme.\cite{Onodera,Goto,Nigg,Puri}
As the two-photon drive strength increases, the system enters a stable cat state 
as the result of the bifurcation.
However, the Kerr effect can be observed 
in some limited systems.
Lechner also proposed the construction method of using controlled-NOT (CNOT) gates.
He showed that the number of the CNOT gates required for constructing a single four-qubit 
interaction is six. 
However, in general, the CNOT gates are complicated to build, 
and it is difficult to use many CNOT gates for constructing the quantum annealing process.
Herein, we propose a method that enables every qubit system interacting 
with nearest-neighbor Ising interactions to realize the LHZ Hamiltonian. 
We dynamically form the many-body interactions by using appropriate pulse sequences.
 
We propose a more concrete method to 
construct the four-qubit interactions without directly using the CNOT gates. 
It is shown that the dynamic pulse sequences by single-qubit rotations and 
two-body interactions enable the formation of the four-body 
and three-body interactions.

The rest of this paper is organized as follows.
In Section~\ref{sec:formation}, we show our dynamical pulse sequence using 
the effective Hamiltonian method in~\cite{tanamoto3}. 
In Section~\ref{sec:results}, we show the numerical results 
of the success probability of our method.
In Section~\ref{sec:discussion}, we discuss our results.
In Section~\ref{sec:conclusion}, we summarize and conclude this study.

%%%%%%%%%%%%%%%%%%%%%%%%%%%%%%%%%%%%  Fig.1
\begin{figure}
\centering
\includegraphics[width=8.6cm]{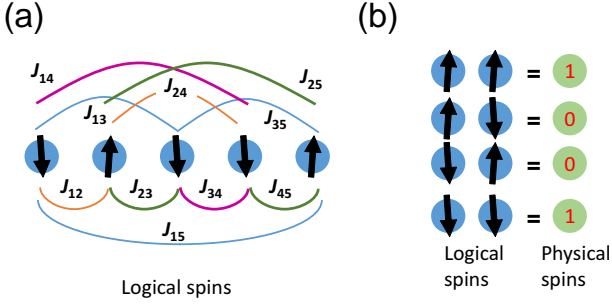}
\caption{LHZ scheme for generating the all-to-all connections.\cite{LHZ}
(a) Ten connections exist for five logical spins.
(b) In LHZ,\cite{LHZ} 
one physical qubit represents two physical spins
such that two parallel spins correspond to qubit "0" and 
two antiparallel spins correspond to qubit "1."
}
\label{fig1}
\end{figure}
%%%%%%%%%%%%%%%%%%%%%%%%%%%%%%%%%%%%%
%%%%%%%%%%%%%%%%%%%%%%%%%%%%%%%%%%%%  Fig.2
\begin{figure}
\centering
\includegraphics[width=8.6cm]{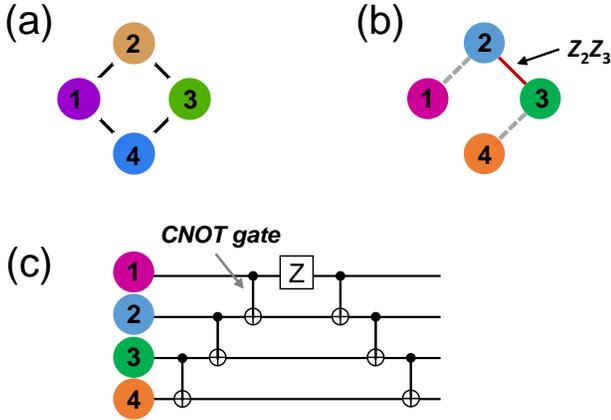}
\caption{
(a) Four-body interaction $Z_1Z_2Z_3Z_4$ in LHZ. 
(b) Realization of the four-body interaction 
by two-body interactions $J_{12}$,$J_{13}$, and $J_{34}$.
The solid brown line shows an initial Hamiltonian that 
uses Eq. (\ref{drot2}). The dotted lines show the required interactions 
to form the four-body interaction.
(c) Lechner's method to construct the four-body interaction.\cite{Lechner1}
}
\label{fig2}
\end{figure}
%%%%%%%%%%%%%%%%%%%%%%%%%%%%%%%%%%%%%

\section{Effective Hamiltonian method}\label{sec:formation}
\subsection{Previously proposed method for constructing many-body interactions}
In the LHZ method, the all-to-all connections 
shown in Fig. 1(a) are realized by the 
replacement of Fig. 1(b).\cite{LHZ}
The key point is to introduce the four-body interaction 
into the Hamiltonian described by
\begin{eqnarray}
\lefteqn{ H^{\rm LHZ}=\sum_{i}[A(t_a)X_i +B(t_a) J_i Z_i]}
\nonumber \\   
&-&\lambda \sum_{i} Z_{C_u^{(i)}} Z_{C_d^{(i)}} Z_{C_l^{(i)}} Z_{C_r^{(i)}},
\label{LHZ}
\end{eqnarray}
where  $A = 1$ and $B = 0$ at $t_a=0$ and A = 0 and B = 1 at the end of the
calculation. $C_c^{(i)}, \{c\in u,d,l,r\}$ are the neighboring qubits of qubit $i$.
The four-body interaction represents the constraint of Fig. 1(b), 
which means that the physical spin states consist of an even number of 
spins. At the boundary sites, this four-body interaction changes to three-body interaction.
Lechner also used the quantum approximate optimization algorithm (QAOA) scheme,
in which the Hamiltonian $H_0+H_{\rm int}$ is separated into components of
the single-qubit rotations $H_0$ and the interaction parts $H_{\rm int}$.\cite{Lechner1,QAOA} 
The rotation angles of the single qubits and the interactions 
are determined by a feedback loop of measuring the outcome of the previous measurements.
Once the interaction part $\lambda \sum_{i} Z_{C_u^{(i)}} Z_{C_d^{(i)}} Z_{C_l^{(i)}} Z_{C_r^{(i)}}$ 
is separated, this part is constructed by a series of qubit operations.

Let us estimate the number of processes required to 
construct the four-body interactions proposed by Lechner.\cite{Lechner1}
For the Ising Hamiltonian $H_{\rm int}=\sum_{i<j}J_{ij}Z_iZ_j$,
the conditional phase flip (CPF) gate is given by
$R_1^z(\theta_4)R_2^z(\theta_4)e^{i\theta_4 Z_1Z_2}$,
where $\theta_4\equiv \pi/4$, and $R_i^\alpha(\theta)\equiv \exp\{ i \theta \alpha_i \}$
is a single-qubit rotation.
The CNOT gate between qubits 1 and 2 is given by 
$U^{\rm CNOT}_{12}=R_2^y(-\theta_4) U_{12}^{\rm CPF}R_2^y(\theta_4)$,
and the time to obtain the CNOT gate 
is given by $\tau_{\rm CNOT}\approx 4 \tau_{\rm rot} +  \tau_{\rm J}$,
where $\tau_{\rm rot}$ represents the time of the single-qubit rotation and $\tau_{\rm J} =\pi/(4J)$.
Then, the time required to obtain the conditions in Fig.~\ref{fig2}(c) is 
given by $ 25 \tau_{\rm rot} +  6\tau_{\rm J}$.
For the XY model,\cite{Schuch} the CNOT gate is expressed by
\begin{eqnarray}
U_{12}^{\rm CNOT}&=&R_1^z(-\theta_4)R_2^x(\theta_4)R_2^z(\theta_4) U_{12}^{\rm iSWAP}
R_1^x(\theta_4) \nonumber \\
&\times & U_{12}^{\rm iSWAP} R_2^z(\theta_4)
\end{eqnarray}
where $U_{12}^{\rm iSWAP}\equiv \exp\{ i\theta_4[X_1X_2+Y_1Y_2]\}$.
Thus, we have $\tau_{\rm CNOT}\approx 4 \tau_{\rm rot} + 2 \tau_{\rm J}$, 
and the time required to obtain the conditions in Fig.~\ref{fig2}(c) is 
given by $ 25 \tau_{\rm rot} +  12 \tau_{\rm J}$.
Thus a lot of qubit operations are required to construct a single four-body interaction.

%%%%%%%%%%%%%%%%%%%%%%%%%%%%%%%%%%%%%%%%%
%%%%%%%%%%%%%%%%%%%%%%%%%%%%%%%%%%%%%%%%%
\subsection{A creation of the four-body interaction using the effective Hamiltonian method}
Here, we show our scheme for creating the 
effective Hamiltonian of Eq. (\ref{LHZ}).
In order to see the true effect of our method, we 
do not use the feedback of the QAOA approach, 
and we simply approximate the time evolution of the total Hamiltonian 
into small intervals of time. 
A given time $t$ is separated into smaller pieces 
$t=\sum_{i=0}^{N_a} \Delta t$ with $\Delta t=t/N_a$ with an integer $N_a$. 
Thus, in our method, the time evolution is expressed by
\begin{equation}
U(t) =\prod_{l=1}^{N_a} U_{\rm unit} (t_l,t_{l-1})
\label{procedure}
\end{equation}
where $t_{N_a}=t$ and $t_0=0$, and $t_l-t_{l-1}=\tau_{sq}+\tau_{mb}$.
\begin{eqnarray}
U_{\rm unit}(t_l,t_{l-1})
&\approx& 
e^{-i\tau_{sq} \sum_i [ A(t_l)X_i +B(t_l) h_i Z_i] }
\nonumber \\
&\times &e^{-iJ\tau_{mb} \sum Z_iZ_jZ_kZ_l}.
\label{tauB}
\end{eqnarray}
By using the Baker-Campbell-Hausdorff(BCH) formula 
$e^{H_1} e^{H_2}=e^{H_3}$, where $H_3={H_1}+{H_2}+[{H_1},{H_2}]/2+[{H_1},[{H_1},{H_2}]]/12-[{H_2},[{H_1},{H_2}]]/12..$, 
we neglect the commutation relations $[{H_1},{H_2}]$.
In our method, the magnitude of the constraint term is adjusted by the 
time period of $\tau_{mb}$.
Once the unitary evolution is separated into 
each component, we can multiply those unitary operations directly one-by-one,
and we can construct many-body interactions 
starting from two-body interactions.
Here we focus on the case of the Ising interaction, 
and we consider the conversion of the two-body interaction $\sum_{i,j} J_{ij}Z_i Z_j$
into the four-body interaction $\sum_{i,j,k,l} J_{ijkl}Z_i Z_j Z_k Z_l$.
The core idea is to apply the effective Hamiltonian method\cite{tanamoto3}
to the Hamiltonian $H_{\rm eff}=\sum Z_iZ_jZ_kZ_l$. 
The effective Hamiltonian $H_{\rm eff}$ is produced 
from its initial form $H_{\rm ini}$
by applying a series of operations
$H^{\rm op}_j$ such that
\begin{equation}
H_{\rm eff} \rightarrow \prod_{j=1}^n \exp(-i \tau^{\rm op}_j H^{\rm op}_j) H_{\rm ini} 
\prod_{j=n}^1 \exp(i \tau^{\rm op}_j H^{\rm op}_j).
\end{equation}
The increase in the degree of the many-body interactions
is carried out by the basic equations:\cite{tanamoto3}
\begin{eqnarray}
e^{-i\theta Z_{1}Z_{2} } X_{1} e^{i\theta Z_{1}Z_{2}} 
&\!=\!& \cos (2\theta) X_{1} +\sin (2\theta) Y_{1} Z_{2},
\label{ZZa}\\
e^{-i\theta Z_{1}Z_{2} } Y_{1} e^{i\theta Z_{1}Z_{2}} 
&\!=\!& \cos (2\theta) Y_{1} -\sin (2\theta) X_{1} Z_{2}\:.
\label{ZZb}
\end{eqnarray}
For example, if we apply the pulse during $\theta=J\tau_J$, we obtain
\begin{eqnarray}
%X_1 &\rightarrow -Y_1 Z_2, \label{drot1}\\
Y_1 &\rightarrow X_1 Z_2 \label{drot2}. %\\
%Z_1 &\rightarrow -Z_1.\label{drot3}
\end{eqnarray}
Repetitions of these equations enable the transformation of $m$-body interactions into ($m+1$)-body interactions.
%The time required to obtain this step given by $\tau_{\rm J}$ 
%can be estimated.

%----------------------------------------------------
% Heisenberg
%----------------------------------------------------
% The CNOT gate is given by~\cite{Burkard} 
% $U^{\rm CNOT}_{12}=R_2^y(-\theta_4) U_{12}^{\rm CPF}R_2^y(\theta_4)$,
% where the conditional phase flip (CPF) gate is expressed by
% \begin{equation}
% U_{12}^{\rm CPF}
% =e^{-i \frac{\pi}{2}} R_1^z(\theta_4) R_2^z(-\theta_4)
% U_{12}^{\sqrt{\rm SWAP}} R_2^z(-2\theta_4) U_{12}^{\sqrt{\rm SWAP}}, 
% \end{equation}
% where $U_{12}^{\sqrt{\rm SWAP}}\equiv \exp\{ i(\pi/8)[X_1X_2+Y_1Y_2+Z_1Z_2]\}$.
% In Fig.X, the six CNOT gates with $Z$-rotation requires 
% $31\tau_{\rm rot}+12\tau_{\sqrt{\rm SWAP}}$.

%%%%%%%%%%%%%%%%%%%%%%%%%%%%%%%%%%%% Fig.3
\begin{figure}
\centering
\includegraphics[width=7cm]{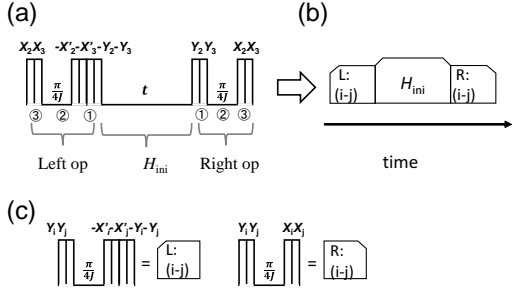}
\caption{
(a) Basic pulse sequence to produce the four-body interaction from the initial 
Hamiltonian $H_{\rm ini}=Z_2Z_3$.
(b) Graphical description of the formation of the four-body interaction 
using the pulse element of (c); see Eq. (\ref{sequence}).
}
\label{fig_pulse}
\end{figure}
%%%%%%%%%%%%%%%%%%%%%%%%%%%%%%%%%%%%%
%%%%%%%%%%%%%%%%%%%%%%%%%%%%%%%%%%%% Fig.4
\begin{figure}
\centering
\includegraphics[width=8.6cm]{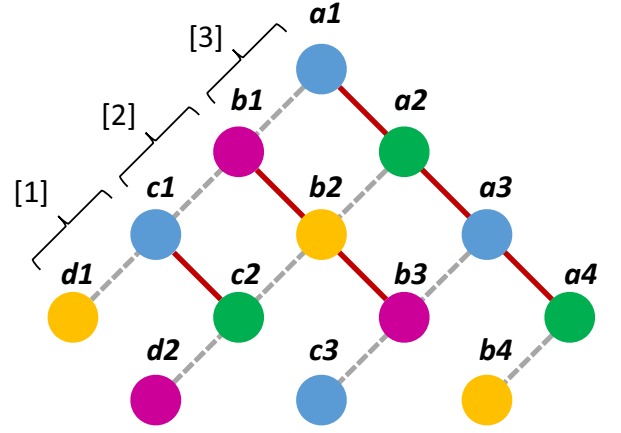}
\caption{
Distribution of interactions to realize all-to-all connections
for six logical qubits by forming the four-body interactions;
an application of our method to the LHZ scheme.
Bold lines show the interactions of the initial Hamiltonian $H_{\rm ini}$.
Dotted lines show the interactions to be formed from the pulse sequence.
Ising interactions are assumed to be switched on and off between the qubits.
}
\label{fig4}
\end{figure}
%%%%%%%%%%%%%%%%%%%%%%%%%%%%%%%%%%%%
%%%%%%%%%%%%%%%%%%%%%%%%%%%%%%%%%%%% Fig.5
\begin{figure}
\centering
\includegraphics[width=8.0cm]{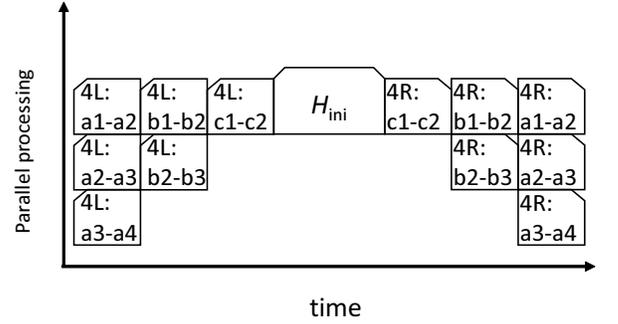}
\caption{
Graphical description of the generation of four-body interaction in the 13-qubit system shown in Fig.~\ref{fig4}.
Parallel processing is possible; see Eq. (\ref{eq13qubits}).}
\label{fig5}
\end{figure}
%%%%%%%%%%%%%%%%%%%%%%%%%%%%%%%%%%%%

%%%%%%%%%%%%%%%%%%%%%%%%%%%%%%%%%%%%%%%%%%%%%%%%%%%%%%%%%%%%%%%%%
%%%%%%%%%%%%%%%%%%%%%%%%%%%%%%%%%%%%%%%%%%%%%%%%%%%%%%%%%%%%%%%%%
As a simple example, we consider the construction of 
a single four-body interaction including four spins (Fig.~\ref{fig2}).
We assume that there is a mechanism for switching interactions on and off.
The initial Hamiltonian is given by $H_{\rm ini}=JZ_2Z_3$, 
where other interactions $Z_1Z_2$, $Z_3Z_4$, and $Z_4Z_1$ are initially switched off.
Once we prepare $H_{\rm ini}=JZ_2Z_3$, we can change this 
Hamiltonian by three steps given by
\begin{eqnarray}
H_{\rm int} &=& 
JZ_2Z_3 \Rightarrow  JX_2X_3 
\ \ \ :[{\rm step1}]\nonumber \\
&\Rightarrow &J(Y_2Z_1)(Y_3Z_4)
\ \ \ :[{\rm step2}]\nonumber \\
&\Rightarrow & JZ_1Z_2Z_3Z_4  \ \ \ :[{\rm step3}] 
\label{sample}
\end{eqnarray}
Here, in step 1, 
we apply a $\pi/2$ pulse around the $y$-axis, 
given by $e^{-i (\pi/4) Y_i}Z_ie^{i (\pi/4) Y_i}=X_i$
for qubits 2 and 3.
In step 2, free running of the system during the period of $\pi/(4J)$
leads to the use of Eq.~(\ref{drot2}). 
In step 3, we apply a $\pi/2$ pulse around the $x$-axis such that 
$e^{i (\pi/4) X_i}Z_ie^{-i (\pi/4) X_i}=Y_i$ for qubits 2 and 3.
These processes are described in Fig.~\ref{fig_pulse}.
\begin{eqnarray}
\lefteqn{ R^X_{2,3}(-\theta_4)
[R^X_{2,3}(2\theta_4) e^{-i\tau_J[Z_1Z_2+Z_3Z_4]}R^X_{2,3}(-2\theta_4)]
}
\nonumber \\
&\times &
R^Y_{2,3}(-\theta_4)
e^{-itZ_2Z_3}
R^Y_{2,3}(\theta_4)
e^{-i\tau_J[Z_1Z_2+Z_3Z_4]}
R^X_{2,3}(\theta_4), \nonumber \\
\label{sequence}
\end{eqnarray}
where $R^\alpha_{2,3}(\theta)=\exp(i\theta [\alpha_2+\alpha_3])$ $(\alpha=X,Y)$.
The square bracket is required to change 
$e^{-i\tau_J[Z_1Z_2+Z_3Z_4]}$ into $e^{i\tau_J[Z_1Z_2+Z_3Z_4]}$,
and we can reduce 
$R^X_{2,3}(-\theta_4)R^X_{2,3}(2\theta_4)=R^X_{2,3}(\theta_4)$ in the first line of the 
equation.
Thus, the required time 
is $5\tau_{\rm rot}+2\tau_J$, which  is 1/6 times less than that of Lechner's method.

%%%%%%%%%%%%%%%%%%%%%%%%%%%%%%%%%%%%%%%%%
The general case is the repetition of the single-four qubit case.
Figure~\ref{fig4} shows the order of the operations for 13 qubits. 
The initial Hamiltonian is given by
\begin{eqnarray}
H_{\rm ini} &=&
Z_{a1}Z_{a2} + 
Z_{a2}Z_{a3} +
Z_{a3}Z_{a4}  \nonumber  \\
&+& 
Z_{b1}Z_{b2} + 
Z_{b2}Z_{b3} +
Z_{c1}Z_{c2}.
\end{eqnarray}
%%%%%%%%%%%%%%%%%%%%%%%%%%
We start from block [1], and blocks [2] and [3] are followed serially.
The detailed pulse sequence of the 13 qubits is 
given by the following, where the bold characters show the operations at each step:
\begin{eqnarray}
\lefteqn{ H_{\rm ini}
%%%%%%%%%%%%%%%%%%[9]
\Rw
Z_{a1}Z_{a2} + 
Z_{a2}Z_{a3} +
Z_{a3}Z_{a4}  }\nonumber  \\
&\!\!+\!\!& 
Z_{b1}Z_{b2} + 
Z_{b2}Z_{b3} +
{\bf X}_{c1}{\bf X}_{c2}
\ {\rm :[step1 ]}    \nonumber    \\
%%%%%%%%%%%%%%%%%%[8]
&\!\!\Rw \!\!&
Z_{a1}Z_{a2} + 
Z_{a2}Z_{a3} +
Z_{a3}Z_{a4}  \nonumber  \\
&\!\!+\!\!& 
Z_{b1}Z_{b2} + 
Z_{b2}Z_{b3} +
{\bf Y}_{c1}{\bf Z}_{d1}{\bf Y}_{c2}{\bf Z}_{d2}    
\  {\rm :[step2]}  
%{\rm apply}\ Z_{c1}Z_{d1}+Z_{c2}Z_{d2}
\nonumber  \\  
%%%%%%%%%%%%%%%%%%[7]
&\!\!\Rw \!\!&
Z_{a1}Z_{a2} + 
Z_{a2}Z_{a3} +
Z_{a3}Z_{a4}  \nonumber  \\
&\!\!+\!\!& 
{\bf X}_{b1}{\bf X}_{b2} + 
{\bf X}_{b2}{\bf X}_{b3} +
Z_{c1}Z_{d1}Z_{c1}Z_{d2}   \ {\rm :[step3 ]}  \nonumber \\
%%%%%%%%%%%%%%%%%%[6]
\!&\!\!\Rw \!\!&\!
Z_{a1}Z_{a2} \!+\! 
Z_{a2}Z_{a3} \!+\!
Z_{a3}Z_{a4}  \nonumber  \\
\!&\!\!+\!\!& \!
{\bf Y}_{b1}{\bf Z}_{c1}{\bf Y}_{b2}{\bf Z}_{c2}  \!+\! 
{\bf Y}_{b2}{\bf Z}_{c2}{\bf Y}_{b3}{\bf Z}_{c3}  \!+\!
Z_{c1}Z_{d1}Z_{c1}Z_{d2}     \ {\rm :[step4 ]}  \nonumber  \\  
%%%%%%%%%%%%%%%%%%[5]
\!&\!\!\Rw \!\!&\!
{\bf X}_{a1}{\bf X}_{a2} \!+\! 
{\bf X}_{a2}{\bf X}_{a3} \!+\!
{\bf X}_{a3}{\bf X}_{a4}  \nonumber  \\
\!&\!\!+\!\!&\! 
Z_{b1}Z_{c1}Z_{b2}Z_{c2}  \!+\! 
Z_{b2}Z_{c2}Z_{b3}Z_{c3}  \!+\!
Z_{c1}Z_{d1}Z_{c1}Z_{d2}     \ {\rm :[step5]}  \nonumber  \\  
%%%%%%%%%%%%%%%%%% 
\!&\!\!\Rw \!\!&\!
{\bf Y}_{a1}{\bf Z}_{b1}{\bf Y}_{a2}{\bf Z}_{b2} \!+\! 
{\bf Y}_{a2}{\bf Z}_{b2}{\bf Y}_{a3}{\bf Z}_{b3} \!+\!
{\bf Y}_{a3}{\bf Z}_{b3}{\bf Y}_{a4}{\bf Z}_{b4}  \nonumber  \\
\!&\!\!+\!\!& \!
Z_{b1}Z_{c1}Z_{b2}Z_{c2} \! +\! 
Z_{b2}Z_{c2}Z_{b3}Z_{c3}  \!+\!
Z_{c1}Z_{d1}Z_{c1}Z_{d2}     \ {\rm :[step6]}   
%\nonumber \\
%& &{\rm apply}\ Z_{b1}Z_{c1}+Z_{b2}Z_{c2}+Z_{b3}Z_{c3}+Z_{b4}Z_{c4}
\nonumber  \\ 
 %%%%%%%%%%%%%%%%%%%%%%%%%%%%%%%%%%%%[3]
&\!\!\Rw \!\!&\!
{\bf Z}_{a1}{\bf Z}_{b1}{\bf Z}_{a2}{\bf Z}_{b2} \!+\! 
{\bf Z}_{a2}{\bf Z}_{b2}{\bf Z}_{a3}{\bf Z}_{b3} \!+\!
{\bf Z}_{a3}{\bf Z}_{b3}{\bf Z}_{a4}{\bf Z}_{b4}  \nonumber  \\
&\!\!+\!\!& 
Z_{b1}Z_{c1}Z_{b2}Z_{c2}  \!+\! 
Z_{b2}Z_{c2}Z_{b3}Z_{c3}  \!+\!
Z_{c1}Z_{d1}Z_{c1}Z_{d2}    \ {\rm :[step7]}   
\nonumber \\
\!\!\!\!\! \label{eq13qubits}
%%%%%%%%%%%%%%%%%%
\end{eqnarray}
Note that the process of $Y_iZ_j\rw Z_iZ_j$, which is the third step in Eq. (\ref{sample}), 
can overlap the next four-body generation step.
Thus, for the three blocks ($N_b=3$), we have $2\times 2$ + 3 = 7 steps of operations.

These processes are easily extended to a general case.
The addition of one block line adds two steps.
As shown in Fig.~\ref{fig4}, the $[N_b]$-th block includes $N_b$ squares and $N_b$ four-qubit interactions.
Thus, the $[N_b]$ block system includes $(N_b+1)(N_b+2)/2+N_b$ qubits 
and $N_b(N_b+1)/2$ initial interactions by $2N_b+1$ steps.
The number of logical qubits, $N_b(N_b+1)/2$ interactions, is feasible. 
The generation time is estimated from the graphical description of Fig.~\ref{fig5}.
The right part includes a time of $N_b (\tau_J+\tau_{\rm rot}) +\tau_{\rm rot}$,
and the left part includes a time of $N_b (\tau_J+2\tau_{\rm rot}) +\tau_{\rm rot}$.
Thus, we need a total time of $N_b (2\tau_J+3\tau_{\rm rot}) +2\tau_{\rm rot}$ by using parallel processing (Fig.~\ref{fig4}). 
%-----------------------
% N of step  
%[1]% 4 qubits= 1 flux => 3steps
%[2]% 8(+4) qubits= 3 flux(+2) => 5steps
%[3]%13(+5) qubits= 6 flux(+3) => 7steps
%[4]%19(+6) qubits= 10flux(+4) => 9steps
%[5]%26(+7) qubits= 15flux(+5) => 11steps
%[n]%(n+1)(n+2)/2+n qubits  ::  sum_i=1(i)=i(i+1)/2  ::  2(i-1)+3 steps :: i-th blocks  
%%%%%%%%%%%%%%%%%%%%%%%%%%%%%%%%%%%%%
%-------------------------------
%  1eV=2.41799×10^14Hz
%  1meV=2.478x10^11Hz= 250GHz
%  1mev=2.41796*10e11[Hz]=2.41796*10e2[GHz]=1.1604*10e2 [K].
%   t=pi/(4J)=3.14/4/2.478*10-11s=3.17ps
%  1uev=2.41796*10e8[Hz]=0.241796[GHz]=1.1604*10e-2 [K]
%    t_J=pi/(4J)=3.17ns   t_rot=100ps
%%%%%%%%%%%%%%%%%%%%%%%%%%%%%%%%%%%%%%%%%%%%%%%%%

%%%%%%%%%%%%%%%%%%%%%%%%%%%%%%%%%%%%%%%%%%%%%%%%%
\subsection{Creation of the three-body interaction}
\begin{figure}
\centering
\includegraphics[width=4.3cm]{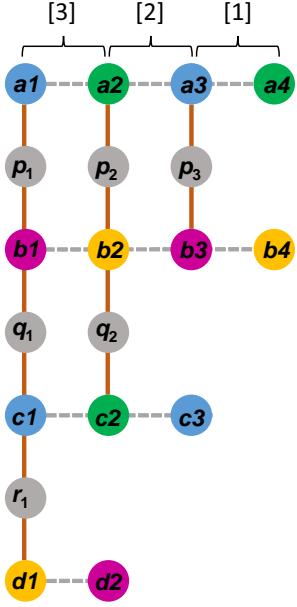}
\caption{
An application of our method to the LHZ scheme of three-body interactions.
The distribution of interactions realizes all-to-all connections
for six logical qubits by three-body interactions.
Bold lines show the interactions of the initial Hamiltonian $H_{\rm ini}$.
Dotted lines show the interactions to be created from the pulse sequence.
The three-body interactions are generated by three blocks.
In the first block [1], the three-body interaction 
regarding the rightmost line is generated.
In the second block [2], the three-body interaction 
regarding the middle line is generated; 
in the third block [3], the three-body interaction 
regarding the left line is generated.
In total, a seven-pulse sequence is required.
The extension to more qubits is straightforward.
}
\label{fig6}
\end{figure}

\begin{figure}
\centering
\includegraphics[width=8cm]{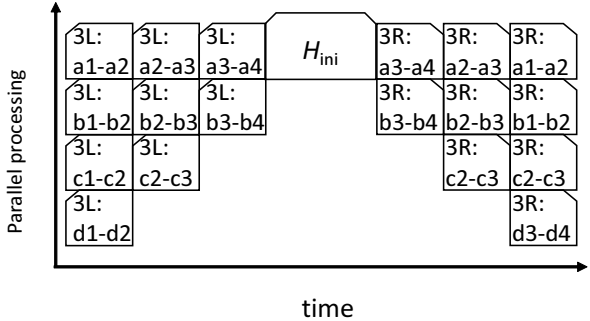}
\caption{
Graphical description of the generation of the three-body interaction of Fig.~\ref{fig6}.
Parallel processing is possible; see Eq. (\ref{3bodyeq}).
}
\label{fig3bodygraph}
\end{figure}

As Lechner\cite{LHZ,Lechner2} derived, 
the LHZ condition is also satisfied by a three-qubit 
interaction using ancilla qubits.
The replacement of the four-body interaction 
by the three-body interaction is expressed by\cite{Lechner2}
\begin{equation}
Z_1Z_2Z_3Z_4 \rw Z_1Z_2 Z_a+Z_3Z_4Z_a,
\end{equation}
where $Z_a$ is the element of the ancilla qubit.
It can be shown that the three-body interaction is derived from the 
two-body interactions similarly to the four-body interaction mentioned above.
Figure~\ref{fig6} shows the formation process of three blocks, 
where six ancilla qubits ($p_i$, $q_i$, $r_i$) are prepared.
The initial Hamiltonian is given by
\begin{eqnarray}
\lefteqn{ 
H_{\rm ini}
%%%%%%%%%%%%%%%%%%[7]
= Z_{a1}Z_{p1} 
+  Z_{a2}Z_{p2} 
+  Z_{a3}Z_{p3} } \nonumber  \\ 
&+&Z_{b1}Z_{p1} 
+  Z_{b2}Z_{p2} 
+  Z_{b3}Z_{p3}  \nonumber  \\
&+&Z_{b1}Z_{q1} 
+  Z_{b2}Z_{q2}  
+  Z_{c1}Z_{q1}    
+  Z_{c2}Z_{q2}   \nonumber  \\
&+&Z_{c1}Z_{r1}
 + Z_{d1}Z_{r1}. 
\end{eqnarray}
We start from block [1], which includes the line with the 
smallest number of qubits.
The transformation of the Hamiltonian is 
carried out stepwise by using Eqs. (\ref{ZZa}) and (\ref{ZZb}), similar to the four-body interaction.
The number of steps is the same as that of the four-body interaction (see Appendix B).
The graphical description of the three-body generation is shown in Fig.~\ref{fig3bodygraph}.
The generation time is the same as that of the four-body interaction 
and is given by $N_b (2\tau_J+3\tau_{\rm rot}) +2\tau_{\rm rot}$.
The difference between the four-body generation and the three-body interaction 
is that the qubits that are controlled are mutually separated in 
the three-body generation case because of the existence of the ancilla qubits.
This will be helpful in fabricating the gate electrodes to control the qubits.
The disadvantage of the three-body interaction array is that 
the number of qubits is larger than that of the four-body interaction case.

%%%%%%%%%%%%%%%%%%%%%%%%%%%%%%%%%%%%%%%%%%%%%%%%%%%%%%%%%%%%%%%%%%%%%%%% Fig.6
%%%%%%%%%%%%%%%%%%%%%%%%%%%%%%%%%%%%
%%%%%%%%%%%%%%%%%%%%%%%%%%%%%%%%%%%%%%%%%%%%%%%%%%%%%%%%%%%%%%%%%%%%%%%% Fig.6
%%%%%%%%%%%%%%%%%%%%%%%%%%%%%%%%%%%%

%%%%%%%%%%%%%%%%%%%%%%%%%%%%%%%%%%%% Fig.cal
\begin{figure}
\centering
\includegraphics[width=5.0cm]{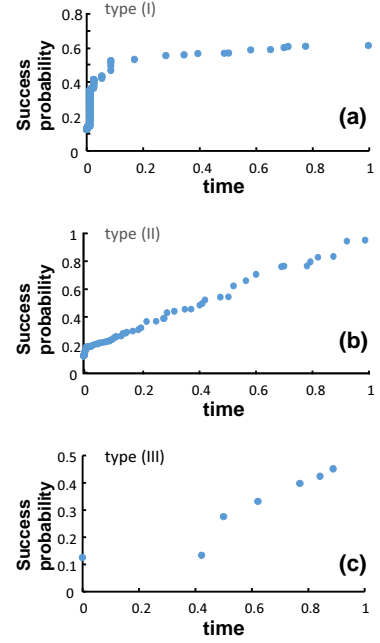}
\caption{
Numerically calculated success probability of the annealing process.
(a) $\Delta(t_a)=1-t_a$,
(b) $\Delta(t_a)=1-\exp(-t_a)$, 
and (c) $\Delta(t_a)=1/\sqrt{t_a+1}$.
$\tau_{sb}=\pi/200$,
$N_{\rm S}=50$ and $J\tau_{\rm M}=\pi/2$.
The annealing time $t_a$ is divided into $N=10^5$ steps.
Each time step has a time interval of $\tau_{\rm S}$+$\tau_{\rm M}$ 
as shown in Eq. (\ref{tauB}) where $\tau_{\rm S}=N_{\rm S}\tau_{sq}$. 
Thus, the real elapsed time 
is estimated by $N(\tau_{\rm S}+\tau_{\rm M}$).
}
\label{figcal}
\end{figure}

\section{Numerical calculation}\label{sec:results}
%%%%%%%%%%%%%%%%%%%%%%%%%%%%%%%%%%%%%%%%%%%%%%
We calculate the success probability of our method 
for the four-body interaction in six qubits.
The time evolution of the unitary matrix is calculated  
using the Chebyshev expansion,\cite{Cheb} 
and overlapping the evolution with the exact wave functions is estimated. 
The initial input data $h_i$ are randomly chosen ($i=1,..,6)$.
The limited number of qubits is caused by the 
calculation resources. For this reason, the number 
of qubits (six qubits) in the three-body interaction is not calculated here.

The three types of the annealing schedules considered for 
$A(t_a)=\Delta(t_a)$ and $B(t_a)=1-\Delta(t_a)$ in Eq. (\ref{LHZ}) are given by
\begin{eqnarray}
\Delta(t_a)&=&1-t_a, \label{deq1} \ \ \ {\rm (I)}  \nonumber \\
\Delta(t_a)&=&1-\exp(-t_a), \label{deq2} \ {\rm (II)}  \nonumber \\
\Delta(t_a)&=&1/\sqrt{t_a+1}. \label{deq3} \ \ \ \ \ {\rm (III)} \nonumber
\end{eqnarray}
In this calculation,  
$A = 1$ and $B = 0$ at $t_a=0$ and $A = 0$ and $B = 1$ at $t_a=1$.
The time $0<t_a<1$ is divided into $N_a$ steps, during each of which 
the single unit of Eq. (\ref{tauB}) is carried out.

In order to use the BCH formula in Eq.(\ref{procedure}), the time step $t_a/N_a$ should be sufficiently small.
When we follow the calculational procedure of Eq.(\ref{procedure}), 
we have to calculate many sets of $U_{\rm unit}(t_l,t_{l-1})$.
In this procedure, the $N_a$ times of the formation of the 
many-body interactions is repeated, and  the operations complexity increases 
as the time step  $t_a/N_a$ becomes smaller.
%The success probability is calculated by the overlap of the unitary operator of our method 
%with that of the exact QA Hamiltonian.
It is found that the success probabilities does not reach one in the calculations 
of the range $N_a\sim 10^6$ (figures not shown).
Thus, we think that, if the time step is sufficiently small, 
we can rearrange the order of the operations Eq.(\ref{procedure}) 
such as 
\begin{eqnarray}
U(t_{l},t_{l-1})U(t_{l+1},t_{l})
&\approx &
e^{-i\tau_{sq} H^{\rm SQ}(t_l)}
e^{-i\tau_{sq} H^{\rm SQ}(t_{l+1})}  \nonumber \\
&\times& e^{-i2 \tau_{mb} H^{\rm MB} },
\end{eqnarray}
where $H^{\rm SQ}(t_l)\equiv \sum_i [ A(t_l)X_i +B(t_l) h_i Z_i]$
and $H^{\rm MB}\equiv J \sum Z_iZ_jZ_kZ_l$.
Then we can collect parts of Eq.(\ref{procedure}),
and the whole unitary operations consist of 
the lumps of smaller processes 
each of  which has $N_{\rm S}$ times of $H^{\rm SQ}$ and $H^{\rm MB}$.
That is, one lump contains
$\Pi_{i=l'}^{l'+N_{\rm S}} e^{-i\tau_{sq}  H^{\rm SQ}(t_i)}$
and
$e^{-i N_{\rm S} \tau_{mb} H^{\rm MB}}$.
This method has the advantage 
of maximizing the effect of the constraints of the four-body interaction,
for $J\tau_{\rm M}=\pi/2+m\pi$ with integer $m$ where $\tau_{\rm M}\equiv N_{\rm S} \tau_{mb}$, 
because of the relationship  
$e^{-i \tau_{M} H^{\rm MB}}
=\cos (J\tau_{M})-i \sin (J\tau_{M})H^{\rm MB}/J$. 
Hereafter, we treat this method to estimate the success probabilities.

\begin{figure}
\centering
\includegraphics[width=5.0cm]{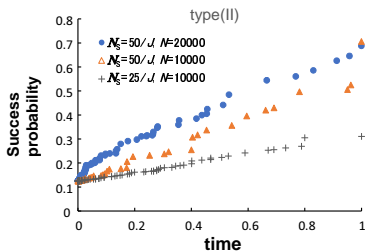}
\caption{
Numerically calculated success probability of the annealing process of type (II) 
 ($\Delta(t_a)=1-\exp(-t_a)$) when the $N$ and $N_{\rm S}$ are reduced.
}
\label{figapp}
\end{figure}

Figure~\ref{figcal} shows the result 
of $N=10^5$ and $N_{\rm S}=50$.
It is found that type (II) is the best for scheduling.
As $N$ and $N_{\rm S}$ become larger,
the success probability increases.
Next we consider whether $N$ can be reduced or not by focusing on the type (II).
Figure~\ref{figapp} shows different parameter regions
of $N$ and $N_{\rm S}$ for the type (II).
It is found that the reduction of $N_{\rm S}$ degrades the success probability.  
Although the $N$ in Fig.~\ref{figapp} is about one-fifth smaller than the $N$ in Fig.~\ref{figcal},
the success probability of Fig.~\ref{figapp} become about 80\% of Fig.~\ref{figcal}.
These results show that the speed to reach to the maximum success probability becomes slower 
when the success probability become close to one.
From the realistic viewpoint, the stopping point of the annealing process 
will depend on the requirement of the accuracy of the individual solution.
%%%%%%%%%%%%%%%%%%%%%%%%%%%%%%%%%%% Fig.cal
%%%%%%%%%%%%%%%%%%%%%%%%%%%%%%%%%%%%%%%%%%%%%%

\section{Discussions}\label{sec:discussion}
We estimate the time required to carry out our processes
regarding the calculations of Fig.~\ref{figcal}.
When we choose $J=100 \mu$eV assuming $\tau_{\rm rot} \ll \tau_{\rm J}$ and $N_b=3$,  
we have $J\tau_{\rm M}=6\tau_{\rm rot}+11\pi/2 \sim 11\pi/2$, and $\tau_{\rm M} \sim 7.15\times 10^{-10}$s.
For $N_{\rm S}=50$, we have $\tau_{\rm S} \sim 4.14\times 10^{-8}$s.
The repetition of $\tau_{\rm M}+\tau_{\rm S}$ by $N=10^{5}$ times 
leads to 
278 $\mu$s as the total annealing time.
Because the operation times are limited by the coherence time of the system, 
we have to reduce the annealing time.
In order to reduce the total annealing time, 
we must increase the strength of the coupling $J$. 
If we apply our idea to a quantum annealing machine based on floating gates (FG)\cite{tanamoto1,tanamoto2} 
with 15 nm width, 100 nm height, and tunneling oxide thickness 3.5 nm, 
we have $J\approx 10.34$ meV and $\tau_{\rm J} \approx 0.304$ ps. 
Then, we have the total annealing time of 2.69 $\mu$s. 
%=100/(2.5837*10^10Hz)=38.7*10-13s=3.87x10-12s.
As $J$ increases, the number of qubits could be increased.
Whether the feedback developed in the QAOA~\cite{QAOA} is effective 
to optimize the number of the annealing process of our model 
is a future problem.

%*-------------------------------------------------------
%* [20200917] 時間はNTB/RR0ということ？100/RR0 つまり2で割った値 50
%*  NTB/RR0*Nx0=50*100000=tJとするとt=5*10^6 /J
%* =5*10^6/2.41799×10^10  [s]=0.0207[s]=200[us]
%-----------------------------------------------------
%*  for ---- J=100ueV=2.41799×10^10Hz
%* 1eV=2.41799×10^14Hz
%*  pi/2/1ueV=3.1415/2/2.41799/10^8=6.49*10^-9 [s]
%* 1K=（絶対温度）	8.61734×10-5eV	=2.08366×10^10Hz
%  J(FG)=124K=1.034*10-2 eV=10.34 meV=2.5837x10^13 Hz
%--------------------------------------------------------
% tau_{\rm M}=11pi/2/100ueV=11pi/2/2.41799×10^10 Hz=7.1459x10^{-10}s
% tau_{\rm S}=100/2/100ueV=50/10^{-4}/2.41799×10^14Hz=20.678x10^{-10}s
%  RR0=2 ということは半分でわっている!!!! NTB=>NTB/2と論文では書こう。
% N*(tau_{\rm M}+tau_{\rm S})=27.824x10^{-5}s=278.24x10^{-6}s
%------------------------------------  FGの場合
% tau_{\rm M}=11pi/2/10.34 meV=11pi/2/10.34/2.41799×10^11 Hz=0.69109x10^{-11}s
% tau_{\rm S}'=100/2/10.34 meV=50/10.34/2.41799×10^11Hz=1.999980.4136x10^{-11}s
% N*(tau_{\rm M}+tau_{\rm S})=2.691x10^{-6}s
%---------------------------------------------------------

%------------------------------------------
% tau_{sb}=pi/2/100=0.01570796

%%%%%%%%%%%%%%%%%%%%%%%%%%%%%%%%%%
\section{Conclusions}\label{sec:conclusion}
In this study, we proposed a method to 
construct four-qubit interactions without directly using CNOT gates in QAMs. We considered concrete pulse sequences for the all-to-all connection of the LHZ method.\cite{LHZ}
We applied the effective Hamiltonian theory\cite{tanamoto3} and  
showed that the form of the four-body interaction can be constructed 
without directly using CNOT gates. 
The processes for generating the four-body interaction and 
the three-body interaction have the same number of steps.
As the number of steps increases, the success probability increases.
The total annealing time is determined by the size of the system 
and the coherence time.
The findings of this study will help reduce computation costs for solving combinatorial problems in quantum annealing. 
We treated the simple case of no feedback in the process of obtaining optimal annealing parameters.
In future work, it should be discussed whether the number of steps can be reduced using the feedback loop as in the QAOA methods.

\begin{acknowledgements}
We are grateful to T. Mori, H. Fuketa,  
J. Deguchi, Y. Nishi,  and H. Goto for the fruitful discussions.
This work was partly supported by MEXT Quantum Leap Flagship Program (MEXT Q-LEAP) Grant Number JPMXS0118069228, Japan.
\end{acknowledgements}

\appendix
\section{Basic formula}
%%%%%%%%%%%%%%%%%%%%%%%%% single-qubit rotations %%%%%5555555555
The operations treated here are derived from the fundamental mathematical 
equations. The single-qubit rotation is given by
\begin{equation}
\exp( -i\theta \sigma^\alpha )\sigma^\beta \exp(i\sigma^\alpha) 
=\cos (2\theta) \sigma^\beta+ \epsilon_{\alpha \beta \gamma}\sin (2\theta)\sigma^\gamma 
\end{equation}
where $\epsilon_{\alpha \beta \gamma}$ is the Levi-Civita symbol ($\{\alpha,\beta,\gamma\}=\{x,y,z\}$),
and $\sigma_\alpha$ are the Pauli matrices. 
These equations are derived by the relationship
$e( i\theta \sigma^\alpha )=\cos \theta + i \sigma_\alpha \sin \theta$.

\section{Steps in the three-body interaction}
The detailed pulse sequence for generating the three-body interactions in 
Fig.~\ref{fig6} is given by the following (bold characters show the operations at
each step):
\begin{eqnarray}
\lefteqn{ H_{\rm ini}
%%%%%%%%%%%%%%%%%%[7]
\Rw 
   Z_{a1}Z_{p1} 
+  Z_{a2}Z_{p2} 
+  {\bf X}_{a3}Z_{p3} } \nonumber  \\ 
&+&Z_{b1}Z_{p1} 
+  Z_{b2}Z_{p2} 
+  {\bf X}_{b3}Z_{p3}  %\nonumber  \\
%&+&Z_{b1}Z_{q1} 
%+  Z_{b2}Z_{q2}  
%+  Z_{c1}Z_{q1}    
%+  Z_{c2}Z_{q2}   \nonumber  \\
%&+&Z_{c1}Z_{r1}
% + Z_{d1}Z_{r1} 
+ ...\ \ {\rm :[step1 ]}\nonumber  \\
%%%%%%%%%%%%%%%%%%
%%%%%%%%%%%%%%%%%%[6]
&\Rw &
   Z_{a1}Z_{p1} 
+  Z_{a2}Z_{p2} 
+  {\bf Y}_{a3}{\bf Z}_{a4}Z_{p3}  \nonumber  \\ 
&+&Z_{b1}Z_{p1} 
+  Z_{b2}Z_{p2} 
+  {\bf Y}_{b3}{\bf Z}_{b4}Z_{p3}      %\nonumber  \\
%&+&Z_{b1}Z_{q1} 
%+  Z_{b2}Z_{q2} 
%+  Z_{c1}Z_{q1}    
%+  Z_{c2}Z_{q2} \nonumber  \\
%&+&Z_{c1}Z_{r1}
%+  Z_{d1}Z_{r1}   
+...\ \ {\rm :[step2 ]}
%{\rm apply} Z_{a3}Z_{a4}+Z_{b3}Z_{b4}
\nonumber \\
%%%%%%%%%%%%%%%%%%[5]
&\Rw &
   Z_{a1}Z_{p1} 
+  {\bf X}_{a2}Z_{p2} 
+  Y_{a3}Z_{a4}Z_{p3}  \nonumber  \\ 
&+&Z_{b1}Z_{p1} 
+  {\bf X}_{b2}Z_{p2} 
+  Y_{b3}Z_{b4}Z_{p3}  \nonumber  \\
&+&Z_{b1}Z_{q1} 
+  {\bf X}_{b2}Z_{q2}  
+  Z_{c1}Z_{q1}    
+  {\bf X}_{c2}Z_{q2}  % \nonumber  \\
%&+&Z_{c1}Z_{r1}
%+  Z_{d1}Z_{r1}  \ \
+... {\rm :[step3 ]}\nonumber \\
%%%%%%%%%%%%%%%%%%[4]
&\Rw &
   Z_{a1}Z_{p1} 
+  {\bf Y}_{a2}{\bf Z}_{a3}Z_{p2} 
+  {\bf Z}_{a3}Z_{a4}Z_{p3}  \nonumber  \\ 
&+&Z_{b1}Z_{p1} 
+  {\bf Y}_{b2}{\bf Z}_{b3}Z_{p2} 
+  {\bf Z}_{b3}Z_{b4}Z_{p3}  \nonumber  \\
&+&Z_{b1}Z_{q1} 
+  {\bf Y}_{b2}{\bf Z}_{b3}Z_{q2} 
+  Z_{c1}Z_{q1}    
+  {\bf Y}_{c2}{\bf Z}_{c3}Z_{q2} % \nonumber  \\
%&+&Z_{c1}Z_{r1} 
%+  Z_{d1}Z_{r1}  
+... \ \ {\rm :[step4 ]}
%{\rm apply} Z_{a2}Z_{a3}+Z_{b2}Z_{b3}+Z_{c2}Z_{c3}
\nonumber \\
%%%%%%%%%%%%%%%%%%[3]
&\Rw &
   {\bf X}_{a1}Z_{p1} 
+  Y_{a2}Z_{a3}Z_{p2} 
+  Z_{a3}Z_{a4}Z_{p3} \nonumber  \\ 
&+&{\bf X}_{b1}Z_{p1} 
+  Y_{b2}Z_{b3}Z_{p2} 
+  Z_{b3}Z_{b4}Z_{p3} \nonumber  \\
&+&{\bf X}_{b1}Z_{q1} 
+  Y_{b2}Z_{b3}Z_{q2}  
+  {\bf X}_{c1}Z_{q1}    
+  Y_{c2}Z_{c3}Z_{q2} \nonumber  \\
&+&{\bf X}_{c1}Z_{r1}
+  {\bf X}_{d1}Z_{r1}   {\rm :[step5]}
\nonumber \\
%%%%%%%%%%%%%%%%%%[2]
&\Rw &
   {\bf Y}_{a1}{\bf Z}_{a2}Z_{p1} 
+  Z_{a2}Z_{a3}Z_{p2} 
+  Z_{a3}Z_{a4}Z_{p3} \nonumber  \\ 
&+&{\bf Y}_{b1}{\bf Z}_{b2}Z_{p1} 
+  Z_{b2}Z_{b3}Z_{p2} 
+  Z_{b3}Z_{b4}Z_{p3} \nonumber  \\
&+&{\bf Y}_{b1}{\bf Z}_{b2}Z_{q1} 
+  Z_{b2}Z_{b3}Z_{q2} 
+  Y_{c1}Z_{c2}Z_{q1}    
+  Z_{c2}Z_{c3}Z_{q2}  \nonumber  \\
&+&{\bf Y}_{c1}{\bf Z}_{c2}Z_{r1}
+  {\bf Y}_{d1}{\bf Z}_{d2}Z_{r1} \ \ {\rm :[step6 ]}
%{\rm apply} Z_{a1}Z_{a2}+Z_{b1}Z_{b2}+Z_{c1}Z_{c2}+Z_{d1}Z_{d2}
\nonumber \\
%%%%%%%%%%%%%%%%%%[1]
&\Rw &
   {\bf Z}_{a1}Z_{a2}Z_{p1} 
+  Z_{a2}Z_{a3}Z_{p2} 
+  Z_{a3}Z_{a4}Z_{p3} \nonumber  \\ 
&+&{\bf Z}_{b1}Z_{b2}Z_{p1} 
+  Z_{b2}Z_{b3}Z_{p2} 
+  Z_{b3}Z_{b4}Z_{p3} \nonumber  \\
&+&{\bf Z}_{b1}Z_{b2}Z_{q1} 
+  Z_{b2}Z_{b3}Z_{q2} 
+  Z_{c1}Z_{c2}Z_{q1}    
+  Z_{c2}Z_{c3}Z_{q2} \nonumber  \\
&+&{\bf Z}_{c1}Z_{c2}Z_{r1}
+  {\bf Z}_{d1}Z_{d2}Z_{r1}\ \ {\rm :[step7 ]}
\label{3bodyeq}
%%%%%%%%%%%%%%%%%%
%%%%%%%%%%%%%%%%%%
\end{eqnarray}

%%%%%%%%%%%%%%%%%%%%%%%%%%%%%%%%%%%%%%%%%%%%%%%%%

\end{document}